# Epitaxially Grown Single-Crystalline SrTiO$_3$ Membranes Using a Solution-Processed, Amorphous SrCa$_2$Al$_2$O$_6$ Sacrificial Layer

Shivasheesh Varshney[1], Martí Ramis[2], Sooho Choo[1], Mariona Coll[2], Bharat Jalan[1]

[1]Department of Chemical Engineering and Materials Science, University of Minnesota, Twin Cities, Minnesota, 55455, USA
[2]ICMAB-CSIC, Campus UAB 08193, Bellaterra, Barcelona, Spain

*Corresponding authors: varsh022@umn.edu; bjalan@umn.edu

**Abstract**

Water-soluble sacrificial layers based on epitaxially-grown, single crystalline $(Ca, Sr, Ba)_3Al_2O_6$ layer are widely used for creating free-standing perovskite oxide membranes. However, obtaining these sacrificial layers with intricate stoichiometry remains a challenge, especially for molecular beam epitaxy (MBE). In this study, we demonstrate the hybrid MBE growth of epitaxial, single crystalline $SrTiO_3$ films using a solution processed, amorphous $SrCa_2Al_2O_6$ sacrificial layer onto $SrTiO_3$ (001) substrates. Prior to the growth, the oxygen plasma exposure was used to first create the crystalline $SrCa_2Al_2O_6$ layer with well-defined surface crystallinity. Utilizing reflection high energy electron diffraction, x-ray diffraction, and atomic force microscopy, we observe an atomic layer-by-layer growth of epitaxial, single crystalline $SrTiO_3$ film on the $SrCa_2Al_2O_6$ layer with atomically smooth surfaces. The $SrCa_2Al_2O_6$ layer was subsequently dissolved in de-ionized water to create free-standing $SrTiO_3$ membranes that were transferred onto a metal-coated Si wafer. Membranes created with Sr-deficiency revealed ferroelectric-like behavior measured using piezo force microscopy whereas stoichiometric films remained paraelectric-like. These findings underscore the viability of using ex-situ deposited amorphous $SrCa_2Al_2O_6$ for epitaxial, single crystalline growth, as well as the importance of point defects in determining the ferroic properties in membranes.

**Introduction**

Free-standing oxide membranes have garnered significant interest due to their usefulness in strain- or strain-gradient engineering, moiré engineering, and artificial heterostructure engineering[1–8]. These capabilities help to tailor the properties such as superconductivity, magnetoresistance, ferroelectricity, metal-insulator transitions, and numerous others[9–17]. A common method to fabricate a free-standing membrane involve the use of chemically etchable sacrificial layers[15–21]. These sacrificial layers enable the release of functional film from the growth substrate through selective etching, without causing damage to the film. This method results in the synthesis of millimeter-sized membranes[15–21]. This technique also promotes the reuse of expensive single crystal substrate following film exfoliation.

Epitaxially-grown, single-crystalline $Sr_3Al_2O_6$, alloyed with $Ca_3Al_2O_6$ and/or $Ba_3Al_2O_6$, has found widespread use as sacrificial layers in pulsed laser deposition (PLD) techniques[9,16,22]. $Sr_3Al_2O_6$ possesses a pseudocubic perovskite structure, with a cubic unit cell having a lattice parameter of 15.844 Å ($\div 4 = 3.961$ Å). The $(Ba, Sr, Ca)_3Al_2O_6$ system allows for continuous adjustment of the lattice parameter within the range of 3.816-4.125 Å, a range frequently observed in typical perovskite oxides (3.8 – 4.1Å). Moreover, the electronegativity of these alkaline-earth metals affects the solubility of sacrificial layer in water, decreasing in the order of increasing electronegativity ($Ba^{2+} < Sr^{2+} < Ca^{2+}$)[23,24]. Among these solid-solutions, $SrCa_2Al_2O_6$, with a lattice parameter of 3.86 Å has been used for perovskite growth in PLD studies[9].

An alternative to PLD, chemical solution deposition (CSD), has also been investigated to create amorphous $SrCa_2Al_2O_6$ layers[23]. CSD is a versatile method to synthesize ternary and quaternary oxides, which can deliver low-cost production in broad applications[25–28]. The method exploits engineering solution-precursor chemistry to achieve $(Sr, Ca)_3Al_2O_6$ composition on single crystal substrates[23,24]. It involves three steps: first, stabilization of precursor solution, then homogeneous deposition by spin coating, followed by thermal annealing to convert precursor gel to phase pure, epitaxial films[23]. In the past, PLD growth of $La_{0.7}Sr_{0.3}MnO_3$ (LSMO) films having a surface roughness of about 2 nm has been explored on a solution-deposited $(Ba, Sr, Ca)_3Al_2O_6$ sacrificial layers[24]. However, the growth of such sacrificial layers with complex stoichiometry is not straightforward in molecular beam epitaxy (MBE). Furthermore, the low vapor pressure and acceptor-like behavior of some elements such as Al (acting as a deep-acceptor in SrTiO$_3$ (STO)) can prohibit their use in oxide MBE. This is evident by the fact that as of 2023, only a small fraction, approximately one-tenth, of all publications (70+) employed MBE for oxide membrane growth[29]. Varshney et al.[29] have recently introduced a novel approach suitable for MBE, involving the use of binary alkaline-earth oxide sacrificial layers. These layers are conducive to MBE processes, as alkaline-earth metals can be sublimated at temperatures ranging from 300-500°C and readily oxidize in molecular oxygen to form the corresponding oxides. Additionally, the binary oxide sacrificial layer (Mg, Ca, Sr, Ba)O provides a wide range of lattice parameter tuning capabilities, and faster dissolution (< 5 minutes)[29]. When combined with complex sacrificial layers, they allow continuous tunning of lattice parameter from 2.98 to 5.12 Å[29].

One approach to incorporate complex sacrificial layers in MBE processes is to prepare them ex-situ and subsequently utilize them as substrates for film growth via MBE. In this study, we grow epitaxial single crystalline STO using hybrid MBE on CSD synthesized, amorphous SrCa$_2$Al$_2$O$_6$ sacrificial layer. We first show recrystallization of air-exposed amorphous SrCa$_2$Al$_2$O$_6$ into a single crystalline, epitaxial film. This recovery of surface crystal structure is possible even after 365+ days of SrCa$_2$Al$_2$O$_6$ sample preparation. We show layer-by-layer growth of STO film with an atomically smooth surface having roughness ~ 0.223 nm. We show exfoliation and transfer of membranes to other substrates followed by structural characterization using X-ray diffraction. The transferred membranes further show an atomically smooth surface with a roughness of 0.268 nm and a bulk-like lattice parameter. We characterize the electrical properties of membranes using piezoforce microscopy (PFM). Our characterization reveals a ferroelectric-like switching and hysteresis loops in Sr-deficient membranes.

**Experiment**

SrCa$_2$Al$_2$O$_6$ composition was prepared on single crystalline STO (001) substrate using CSD method, employing a metal nitrate precursor route as shown in Figure 1a. The precursor solution was prepared in SrCa$_2$Al$_2$O$_6$ stoichiometric amounts of Sr(NO$_3$)$_2$ strontium nitrate (> 99%), Ca(NO$_3$)$_2$ calcium nitrate (> 99%), and Al(NO$_3$)$_3$.9H$_2$O hydrated aluminum nitrate (> 98%), which were dissolved in Milli-Q water with citric acid C$_6$H$_8$O$_7$ (> 99%) to obtain 0.1M solution. This solution was spin coated to obtain 20 nm thick SrCa$_2$Al$_2$O film on STO substrate and was annealed in an oxygen furnace for 1 hour prior to being sealed in air in a plastic bag as shown in Figure 1b-d. The SrCa$_2$Al$_2$O$_6$ samples exposed to air inside the plastic bag for up to ~180, 270, and 365 days were loaded into the load lock of MBE chamber (Scienta Omicron Inc.). The moisture in the load lock that entered during loading of the sample was removed by baking the chamber for 2 hours at 150 ˚C. A base pressure of $3 \times 10^{-9}$ Torr was achieved after lamp heating the load lock. The sample was then transported in-situ to the MBE growth chamber, which has a base pressure of $5 \times 10^{-9}$ Torr. Reflection high-energy electron diffraction (RHEED) (Staib Instruments) was used to determine the surface crystallinity of the as-loaded sample. A RF oxygen plasma source operated at 250 W at oxygen pressure of $8 \times 10^{-6}$ Torr was used for annealing while the sample was ramped up to 950 ˚C (at 25˚C/min until 650 ˚C and 10˚C/min from 650 to 950 ˚C). The sample was annealed at 950 ˚C for 20 min prior to the growth. Four unit-cell of seed layer of STO were grown on SrCa$_2$Al$_2$O$_6$ at 950 ˚C using a hybrid MBE approach[30], by co-deposition of Sr (99.99% Sigma Aldrich), titanium tetraisopropoxide (TTIP) 99.999% Sigma Aldrich, and oxygen plasma. Subsequently, the sample was annealed again under oxygen plasma for 20 min before restarting the growth of the final STO layer. Note the TTIP precursor was supplied using a gas inlet system consisting of a linear leak valve and a Baratron monometer[30]. The Sr beam equivalent pressure (BEP) was fixed at $7.7 \times 10^{-8}$ Torr and oxygen plasma at 250 W and oxygen pressure of $5 \times 10^{-6}$ Torr. Tunning TTIP BEP allowed changing stoichiometry in the STO film. Ex-situ surface characterization using atomic force microscopy (AFM) in peak-force mode was performed to check the surface morphology and high-resolution X-ray diffraction (HRXRD) was performed using Rigaku SmartLab XE thin film diffractometer equipped with Cu K$_α$ radiation. The XRD measurements were used to examine phase purity, crystallinity, and the out-of-plane lattice parameter. The thickness of STO films was measured using grazing incidence X-ray reflectivity (GIXR) with data fitted using GenX software.

For exfoliation, a polydimethylsiloxane (PDMS) supporting layer was applied on top of the STO/ SrCa$_2$Al$_2$O$_6$/ STO (001) sample. The top surface of the heterostructure adheres to the PDMS stamp. The sample is subsequently placed in deionized water at room temperature for 4-5 days to ensure complete dissolution of SrCa$_2$Al$_2$O$_6$ sacrificial layer. After dissolution, STO substrate was detached from PDMS/STO structure. The PDMS stamp with the STO membrane was then applied to the Au-coated Si substrate placed on a hot plate at 120 ˚C for 5 min. PDMS was gently peeled off resulting in the transfer membrane on Au-coated Si substrate. HRXRD and AFM were used to characterize the transferred membrane. Piezoforce microscopy (PFM), Bruker NanoScope Icon, was used to measure the piezo response of the membranes. Pt-coated Ir cantilevers were used for PFM measurements in vertical domain scan mode.

**Results and Discussion**

Figure 1 shows the process employed to obtain crystalline 20 nm $SrCa_2Al_2O_6$ layer on STO (001) substrate. First, the solution processing (CSD) is used to deposit $SrCa_2Al_2O_6$ (Figure 1a). To improve the crystallinity, the as-prepared $SrCa_2Al_2O_6$ / STO (001) undergoes an optimized thermal treatment in a tubular furnace at 900 °C for 30 min with a heating/cooling rate of 25 °C/min and $O_2$ flow 0.6 L/min as shown in Figure 1b. AFM of the as-prepared sample shows a surface roughness of ~ 0.9 nm (Figure 1c). The $SrCa_2Al_2O_6$/STO sample is sealed in air in a plastic bag for storage and transportation to the MBE system (Figure 1d). Figure 1e shows a schematic of the MBE chamber, where $SrCa_2Al_2O_6$/ STO sample is loaded for the STO growth after ~180, 270, and 365 days of its preparation. RHEED image of the as-loaded sample show an amorphous film, which starts to recrystallize as the substrate temperature is increased from 100 ˚C to 950 ˚C in presence of oxygen plasma (Figure S1). As shown in Figure 1f, RHEED image begin to show diffraction pattern at 100 ˚C, indicating crystallization because of exposure of oxygen plasma. As previously reported, RHEED shows clearer diffraction spots at 825 ˚C, indicating removal of surface carbonates layer[31]. A streaky RHEED pattern of $SrCa_2Al_2O_6$ is observed at 950 ˚C, that indicates epitaxial, single crystalline $SrCa_2Al_2O_6$ on STO (001) surface. Plasma annealing removes the surface carbonates formed upon exposure to air and helps to achieve single crystallinity. It should be noted that this recovery is possible even after ~180 (sample 1), 270 (sample 2), and 365 (sample 3) days of $SrCa_2Al_2O_6$ preparation using CSD method, as demonstrated by the RHEED pattern before and after plasma annealing, which shows the transformation of amorphous $SrCa_2Al_2O_6$ into epitaxial films (Figure 1f). These results thus confirm that the high temperature oxygen plasma annealing in MBE can achieve the crystallinity of complex oxides. Although the mechanism of $SrCa_2Al_2O_6$ crystallization into an epitaxial film has not been addressed, this template effectively facilitates epitaxial growth.

Figure 2a shows the RHEED intensity oscillations during STO growth as a function of time during growth indicating atomic layer-by-layer growth mode. The growth rate calculated from the RHEED oscillations is 1.73 nm/min, thus giving a film thickness of 104 nm for 1 hour of growth. The RHEED images shown in inset of Figure 2a show streaky patterns indicating epitaxial film with a smooth surface. It is important to note the RHEED oscillations were observed only when a 4 unit-cell of STO/SrCa$_2$Al$_2$O$_6$/ STO (001) was grown and annealed. No oscillations were observed if STO seed layer was not used (Figure S2a). However, the STO films without seed layer also remain epitaxial and single crystalline, as observed by RHEED (Figure S2b-c). The 2$\theta$-$\omega$ coupled scan shows phase-pure and single crystalline STO film (Figure 2b). The out-of-plane lattice parameter extracted from the (002) peak of STO is 3.905 ± 0.002 Å (identical to the bulk) and from (008) peak of SrCa$_2$Al$_2$O$_6$ is 15.39 ± 0.002 Å (÷4 = 3.849 Å), indicating a stoichiometry of Sr$_{0.9}$Ca$_{2.1}$Al$_2$O$_6$ (assuming relaxed lattice parameter). The bulk-like lattice parameter indicates that the STO film is nearly-stoichiometric. Inset of Figure 2b shows an AFM image revealing the surface is atomically smooth with a surface roughness of 0.223 nm. The RHEED image (inset to Figure 2b) taken after growth confirms epitaxial film. Non-stoichiometric (Sr-deficient) STO films were grown on SrCa$_2$Al$_2$O$_6$ layer under the excess of TTIP. A peak shift is observed in 2$\theta$-$\omega$ coupled scan due to non-stoichiometry indicating a lattice parameter of 3.936 ± 0.002 Å and the RHEED image reveals an additional half order diffraction (Figure S3a-c). Further, the GIXR of the stoichiometric film shows a film thickness of 104 nm (Figure S3c), in agreement with the expected layer thickness that is calculated from RHEED intensity oscillations.

Figure 3a shows the schematic illustration of exfoliation and transfer process combined with structural characterizations. The entire 5 × 5 mm membrane was exfoliated as shown in inset of Figure 3a. An uncracked area (> 100 μm$^2$) was transferred onto Au-coated Si substrate. Figure 3b shows a wide angle 2$\theta$-$\omega$ coupled scan of the transferred membrane revealing the STO film peaks, Au peaks, and Si peaks. No SrCa$_2$Al$_2$O$_6$ peaks were observed. Fine coupled scans (Figure 3c) reveal Sr-deficient membrane with a lattice parameter of 3.912 ± 0.002 Å and a nominally stoichiometric membrane with a lattice parameter of 3.905 ± 0.002 Å. The full-width at half maximum (FWHM) of the rocking curve around the STO (002) peak reveals that FWHM increases to 1.40° in transferred membrane as compared to 0.46° in the as-grown film (Figure S4). This increase could be due to a macroscopic disorder in the transferred membrane, arising from wrinkles and air-bubbles.

Finally, to probe the dielectric response of these STO membranes, we used PFM and performed box-like switching and hysteresis loops measurements. As shown in Figure 4, we first applied -10 V over 6 × 6 μm$^2$ area, then +10 V over 4 × 4 μm$^2$ area. Finally, 7 × 7 μm$^2$ area was scanned with 0 V tip bias. A box-like pattern with 180° phase contrast was obtained on the Sr-deficient membrane (Figure 4a), whereas no phase contrast was observed in stoichiometric sample. A similar contrast was observed in amplitude and phase when scans were done at other places over the same Sr-deficient membrane (Figure S5). Phase and amplitude hysteresis loops were obtained on the Sr-deficient membrane (Figure 4b and S6). These results indicate room-temperature ferroelectric-like behavior in STO and are consistent with properties commonly observed in Sr-deficient STO thin films[32–34]. STO, being an incipient ferroelectric can show ferroelectric properties under strain, or stoichiometry deviations, electric field, and alloying with (Ca, Ba)[35–40].

**Conclusion**

In summary, we have demonstrated that the CSD-grown sacrificial layer of $SrCa_2Al_2O_6$ can be transformed using high temperature, oxygen plasma annealing from an amorphous state to an epitaxial single crystalline film even after 365 days of exposure to air. An atomically smooth, phase pure, epitaxial STO film was subsequently grown on $SrCa_2Al_2O_6$ using hybrid MBE. The sacrificial layer is dissolved to exfoliate millimeter-sized membranes, which are transferred on to Au-coated Si substrate. The membranes show a bulk-like structural properties with an atomically smooth surface. Further, we demonstrate ferroelectric-like behavior in Sr-deficient STO membranes using PFM measurements. Our results highlight a facile method to fabricate membranes using CSD by tackling the challenge to grow complicated stoichiometry of $SrCa_2Al_2O_6$ in MBE, and yet using MBE to grow single crystal functional oxide membranes. Our study broadens the platform of synthesis of sacrificial layers and functional membranes, thus creating new opportunities and applications in science and engineering.

**Figures (color online):**

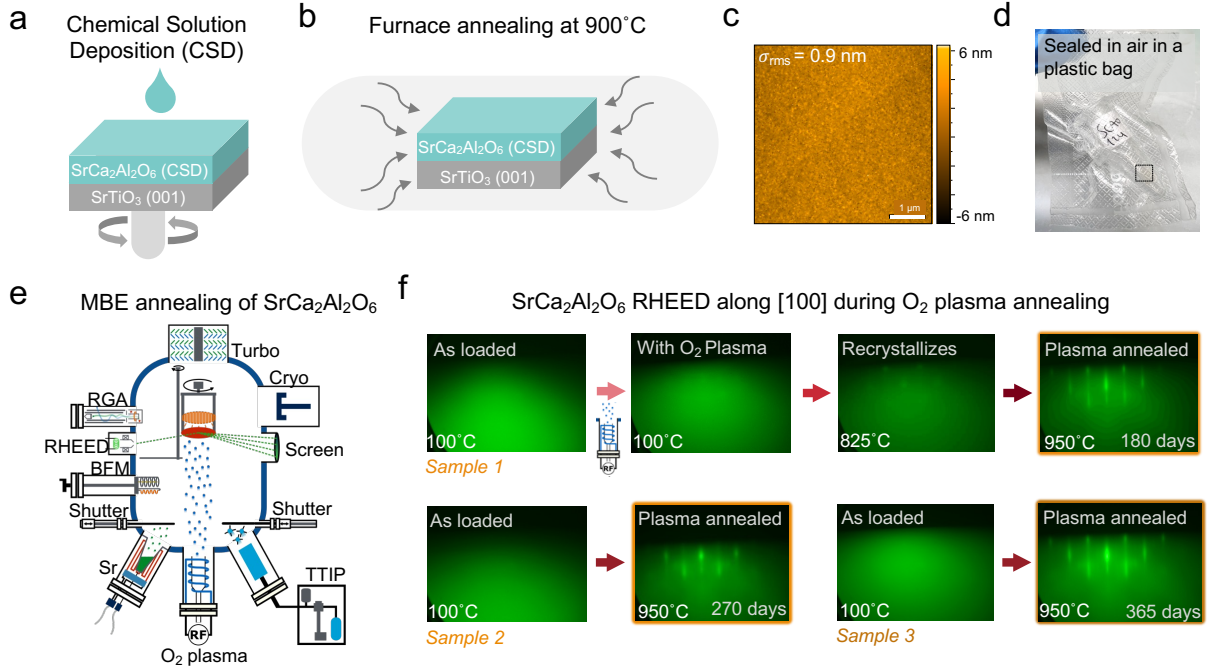

**Figure 1: Synthesis process illustrating a way to obtain epitaxial, single crystalline SrCa$_2$Al$_2$O$_6$ film from a solution-grown amorphous SrCa$_2$Al$_2$O$_6$ layer on SrTiO$_3$ substrate.** (a) Schematic illustrating chemical solution deposition (CSD) method of SrCa$_2$Al$_2$O$_6$ sacrificial on STO (001). (b) The as-prepared SrCa$_2$Al$_2$O$_6$ /STO (001) is oxygen annealed at 900 ˚C. (c) AFM of the SrCa$_2$Al$_2$O$_6$ after furnace annealing showing a smooth surface with rms roughness of ~ 0.9 nm. (d) Picture of the plastic bag, where the as-prepared SrCa$_2$Al$_2$O$_6$ after annealing were sealed in air. (e) Schematic of the hybrid MBE system for vacuum annealing of the as-prepared samples with oxygen (O$_2$) plasma. (f) RHEED evolution during plasma annealing of SrCa$_2$Al$_2$O$_6$ on STO (001) substrate. The as-loaded sample 1 is amorphous at 100 ˚C. As soon as O$_2$ plasma starts, the sample shows spots in RHEED, indicating crystallinity. As temperature increases, SrCa$_2$Al$_2$O$_6$ crystallizes. RHEED image at 950 ˚C shows SrCa$_2$Al$_2$O$_6$ is epitaxial on STO substrate. This recovery of RHEED is possible even after 270 days (Sample 2) and 365 days (Sample 3) of SrCa$_2$Al$_2$O$_6$ sample preparation.

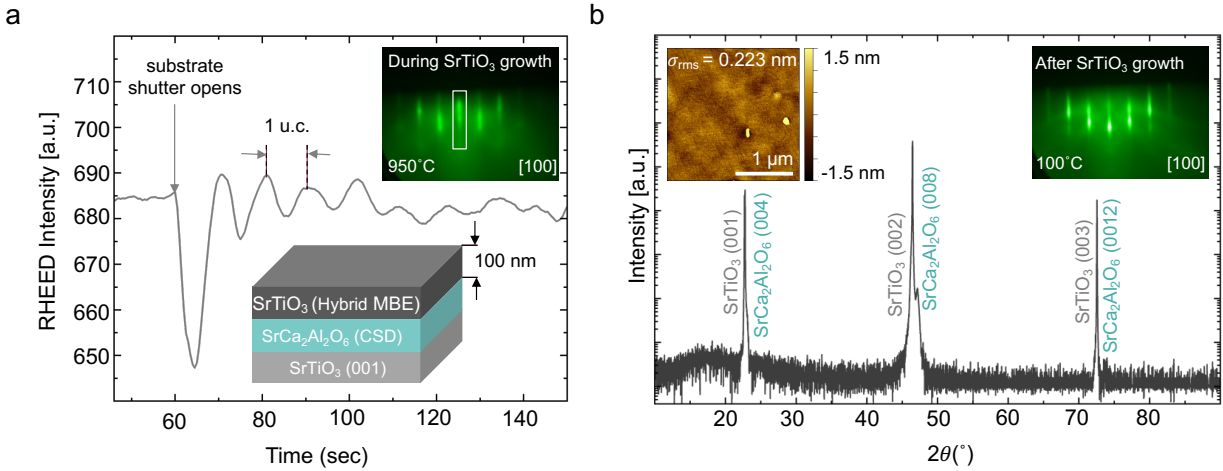

**Figure 2: Epitaxial single-crystalline growth of STO on SrCa$_2$Al$_2$O$_6$/ STO (001) substrate.** (a) RHEED intensity oscillations during STO growth, indicating atomic layer-by-layer growth mode. Inset shows a sample schematic and RHEED image taken during growth taken at 950 ˚C. Intensity profile is shown for the RHEED spot within a box. (b) 2$\theta$-$\omega$ coupled scan of 104 nm STO/ 20 nm SrCa$_2$Al$_2$O$_6$ / STO (001) substrate, indicating single-crystalline, phase-pure, stoichiometric STO film. Inset shows RHEED image after STO growth and the AFM image of STO with rms roughness of 0.223 nm.

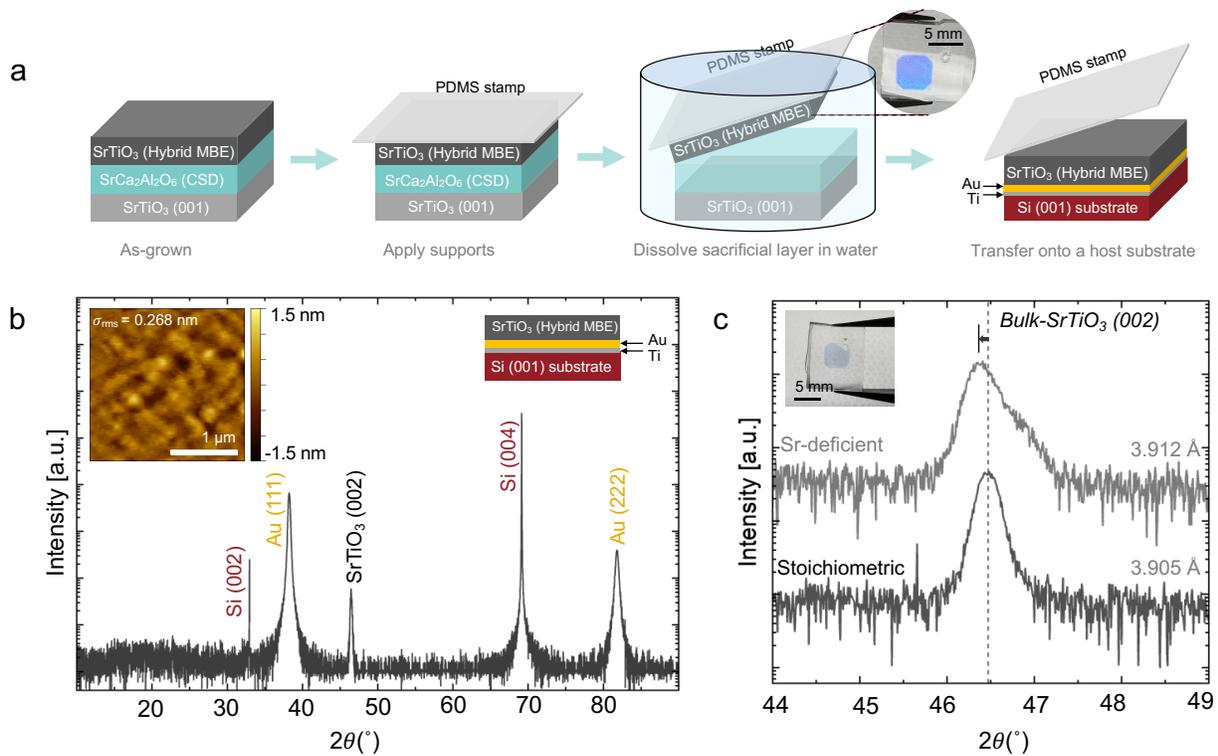

**Figure 3: Exfoliation and transfer of STO membranes to Au-coated Si substrates.** (a) Schematic illustration of the process used for exfoliation and transfer of STO membranes onto a host substrate. Inset shows the stoichiometric membrane on the PDMS stamp. (b) 2θ-ω coupled scan of 104 nm STO on Au-coated Si substrate. The insets show AFM image of transferred membrane, and a sample schematic. (c) Fine coupled 2θ-ω scan around STO (002) peak indicating stoichiometric membrane with bulk-like lattice parameter, 3.905 ± 0.002 Å and a sightly Sr-deficient film with a lattice parameter of 3.912 ± 0.002 Å. Inset shows the Sr-deficient membrane on the PDMS stamp.

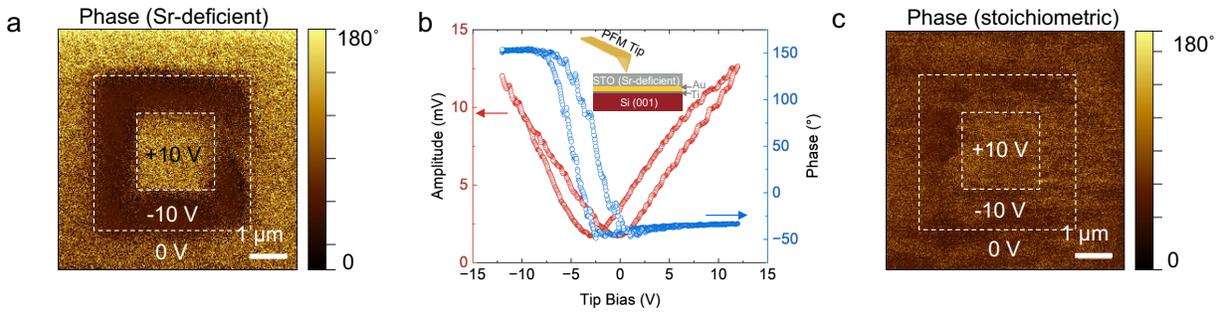

**Figure 4: PFM measurements for Sr-deficient and stoichiometric STO membranes.** (a) Phase contrast on applying ± 10 V d.c. bias to the tip for Sr-deficient sample. (b) PFM phase and amplitude showing hysteretic behavior for Sr-deficient membrane. Inset shows a sample schematic (c) PFM phase with application of ± 10 V d.c. bias to the tip for stoichiometric STO sample showing no contrast.

**Acknowledgements**

Synthesis of membrane and characterization (S.V. and B.J.) were supported by the U.S. Department of Energy through DE-SC0020211. Device characterization (fabrication and dielectric measurements) was supported as part of the Center for Programmable Energy Catalysis, an Energy Frontier Research Center funded by the U.S. Department of Energy, Office of Science, Basic Energy Sciences at the University of Minnesota, under Award No. DE-SC0023464. S.C. acknowledges support from the Air Force Office of Scientific Research (AFOSR) through Grant Nos. FA9550-21-1-0025, FA9550-21-0460. Film growth was performed using instrumentations funded by AFOSR DURIP awards FA9550-18-1-0294 and FA9550-23-1-0085. Parts of this work were carried out at the Characterization Facility, University of Minnesota, which receives partial support from the NSF through the MRSEC program under award DMR-2011401. Exfoliation of films and device fabrication was carried out at the Minnesota Nano Center, which is supported by the NSF through the National Nano Coordinated Infrastructure under award ECCS-2025124. M.C. acknowledges funding by MICIN PID2020-114224RB-I00/AEI/10.13039/501100011033. The work of M.R. has been done in the framework of the doctorate in Materials Science of the Autonomous University of Barcelona.

**Authors contribution**

S.V. and B.J. conceived the idea and designed the experiments. S.V. grew the STO films. M.R. grew the $SrCa_2Al_2O_6$ films under the supervision of M.C. S.V. transferred membranes and performed structural characterization. S.V. and S.C. performed the electrical measurements. S.V., M.C. and B.J. wrote the manuscript. All authors contributed to the discussion and manuscript preparation. B.J. directed the overall aspects of the project.

**Competing Interest Statement:**

There are no conflicts to declare.